\documentclass
[preprint,showpacs,byrevtex,10pt,twocolumn,tightenlines,prl,letterpaper]{revtex4}%
\usepackage{amsfonts}
\usepackage{amsmath}
\usepackage{amssymb}
\usepackage[dvips]{graphicx}
\usepackage{hyphenat}%
\providecommand{\U}[1]{\protect\rule{.1in}{.1in}}

\begin{document}
\preprint{ }
\title{Nonequilibrium transport and Electron-Glass effects in thin Ge$_{\text{x}}$Te films}
\author{Z. Ovadyahu}
\affiliation{Racah Institute of Physics, The Hebrew University, Jerusalem 91904, Israel }


\begin{abstract}
We report on results of nonequilibrium transport measurements made on thin
films of germanium-telluride (Ge$_{\text{x}}$Te) at cryogenic temperatures.
Owing to a rather large deviation from stoichiometry ($\approx$10\% of Ge
vacancies), these films exhibit p-type conductivity with carrier-concentration
\textit{N}$\geq$10$^{\text{20}}$cm$^{\text{-3}}$ and can be made either in the
diffusive or strongly-localized regime by a judicious choice of preparation
and post-treatment conditions. In both regimes the system shows persistent
photoconductivity following excitation by a brief exposure to infrared
radiation. Persistent photoconductivity is also observed in Ge$_{\text{x}}$Te
samples alloyed with Mn. However, in both Ge$_{\text{x}}$Te and
GeMn$_{\text{x}}$Te$_{\text{y}}$ the effect is much weaker than that
observable in GeSb$_{\text{x}}$Te$_{\text{y}}$ alloys suggesting that antimony
plays an important role in the phenomenon. Structural studies of these films
reveal an unusual degree of texture that is rarely realized in
strongly-disordered systems with high carrier-concentrations.
Anderson-localized samples of Ge$_{\text{x}}$Te exhibit non-ergodic transport
which are characteristic of intrinsic electron-glasses, including a well
developed memory-dip and slow relaxation of the excess conductance created in
the excited state. These results support the conjecture that electron-glass
effects with inherently long relaxation times is a generic property of all
Anderson-localized systems with large carrier-concentration.

\end{abstract}
\maketitle

\section{Introduction}

The non-interacting Anderson-insulating phase has been called a Fermi-glass
\cite{1}, presumably inspired by the spatial arrangement of the localized
electronic wavefunctions resembling an amorphous structure. Further
considerations, admitting for the long-range Coulomb interaction, inevitably
present in a medium devoid of metallic screening, led several authors to
suggest that a glassy phase, so called electron-glass (EG), should be
observable in real systems \cite{2,3,4,5,6,7,8,9,10,11}.

Experimental observations consistent with the anticipated glassy behavior were
reported in several Anderson-localized systems \cite{12,13}. On the other
hand, these effects were not seen in either Si or GaAs, systems that are
readily made insulating and exhibit strong-localization transport properties.
To some researchers, this shed doubts on the notion that the electron-glass is
a generic phenomenon peculiar to the Anderson insulating regime.

It has been conjectured that the absence of electron-glass features in Si and
GaAs is related to their relative low carrier-concentration \textit{N}
\cite{14}. This was based on the observation that the dynamics in amorphous
indium-oxide films \cite{15} becomes much faster once \textit{N}$\leq
$10$^{\text{20}}$cm$^{\text{-3}}$. To date, a common feature in \textit{all}
Anderson-insulators that exhibit intrinsic EG effects, \textit{in addition }to
being strongly-localized, is their high carrier-concentrations, typically with
\textit{N}$\geq$10$^{\text{20}}$cm$^{\text{-3}}$. By \textquotedblleft
intrinsic" we mean that the non-ergodic effects appear in a given substance
\textit{independently} of the way the sample was prepared to achieve the
required parameters (resistance at the measuring temperature,
carrier-concentration, and dimensionality determined by the hopping-length to
thickness ratio). Most importantly, the system has to exhibit a memory-dip
with a width that is commensurate with the carrier-concentration of the
material \cite{14}. This distinction is important; slow conductance relaxation
by itself is not necessarily a sign for EG, slow relaxation (and 1/f noise)
may occur in lightly-doped semiconductors, presumably due to extrinsic effects
\cite{13}.

The correlation between high carrier-concentration and sluggish relaxation
rates may suggest the relevance of many-body effects. However, a case may also
be made for the difference in \textit{disorder }being the reason behind the
correlation with carrier-concentration. Note that the requirement of strong
localization means that a system with higher carrier-concentration has
perforce more\textit{ disorder }(required to overcome the higher kinetic
energy associated with higher carrier-concentration). One may then argue that
the reason for the slow relaxation (and the various glassy features) exhibited
by systems with higher Fermi energies is their considerably larger disorder
rather than due to correlation effects. It may transpire that there is a
peculiar \textit{type} of disorder that exists in the high-n systems that
lightly-doped semiconductors cannot sustain and it is this "defect" which
slows down the relaxation of the system from an out-of-equilibrium state. It
is therefore of interest to experimentally test more systems with diversified
structural properties.

This work describes transport measurements on Ge$_{\text{x}}$Te samples, yet
another system with carrier-concentration \cite{16} \textit{N} $\geq
$10$^{\text{20}}$cm$^{\text{-3}}$, somewhat above the empirical limit for
observing electron-glass effects (when the system is strongly-localized).
Comparison with films made with the alloy GeSb$_{\text{x}}$Te$_{\text{y}}$
reveals a \textit{much} weaker persistent-photoconductivity \cite{17} (PPC) in
the Ge$_{\text{x}}$Te films. The relaxation law from the photo-excited state
also differ from that observed in GeSb$_{\text{x}}$Te$_{\text{y}}$ presumably
associated with another kind of a defect. The microstructure of the
Ge$_{\text{x}}$Te films prepared by the method of this work show some unique
features such as preferred orientation (texture) over a extremely large
spatial scale, a single-crystal-like attribute. Yet, Anderson-localized films
of this material exhibit nonequilibrium transport effects including a
memory-dip characteristic of the electron-glass phase just like found in all
previously studied systems. The implications of these finding to the origin of
slow relaxation of electron-glasses are discussed.

\subsection{Sample preparation and characterization}

Samples used in this work were prepared by e-gun depositing GeTe onto room
temperature substrates in a high-vacuum system (base pressure 1$\cdot
$10$^{\text{-7}}$mbar) using rates of 1-2\AA /second. The source material was
99.999\% pure GeTe (Equipment Support Company, USA). Film thickness was in the
range of 30-75\AA . Lateral dimensions of the samples used for the low
temperature studies were 0.3-0.5mm long and 0.5mm wide. Two types of
substrates were used; 1mm-thick microscope glass-slides, and 0.5$\mu$m
SiO$_{\text{2}}$ layer thermally grown on $\langle$100$\rangle$ silicon
wafers. These were boron-doped and had bulk resistivity $\rho\simeq$ 2$\cdot
$10$^{\text{-3}}\Omega$cm, deep into the degenerate regime. This makes this
substrate suitable to perform as a gate-electrode even at low temperatures.
Samples deposited on these wafers were configured as three-terminal devices
for field-effect measurements. These were designed to probe $\partial
$n$/\partial\mu$($E$), the thermodynamic density of states versus energy of
the material as well as to test for electron-glass behavior. Samples prepared
on microscope glass-slides were mainly used for optical characterization and
for Hall-Effect measurements, both performed at room-temperatures.

Each deposition batch included samples for optical excitation measurements,
samples for Hall-effect measurements, and samples for structural and chemical
analysis using a\ transmission electron microscope (TEM). For the latter
study, carbon-coated Cu grids were put close to the sample during its
deposition and received the same post-treatment as the samples used for
transport measurements.

The Philips Tecnai F20 G2) was used to characterize the films composition
(using energy dispersive spectroscopy, EDS) and microstructure. The EDS
typically showed Ge deficiency so, we refer to our deposited films as
Ge$_{\text{x}}$Te. The Cary-1 spectrophotometer was used for optical measurements.

Films deposited at room temperatures were amorphous. TEM and associated
diffraction pattern of typical Ge$_{\text{x}}$Te sample deposited an hour
prior to being inserted to the TEM are shown in Fig.1.%
\begin{figure}[ptb]%
\centering
\includegraphics[
height=3.039in,
width=3.039in
]%
{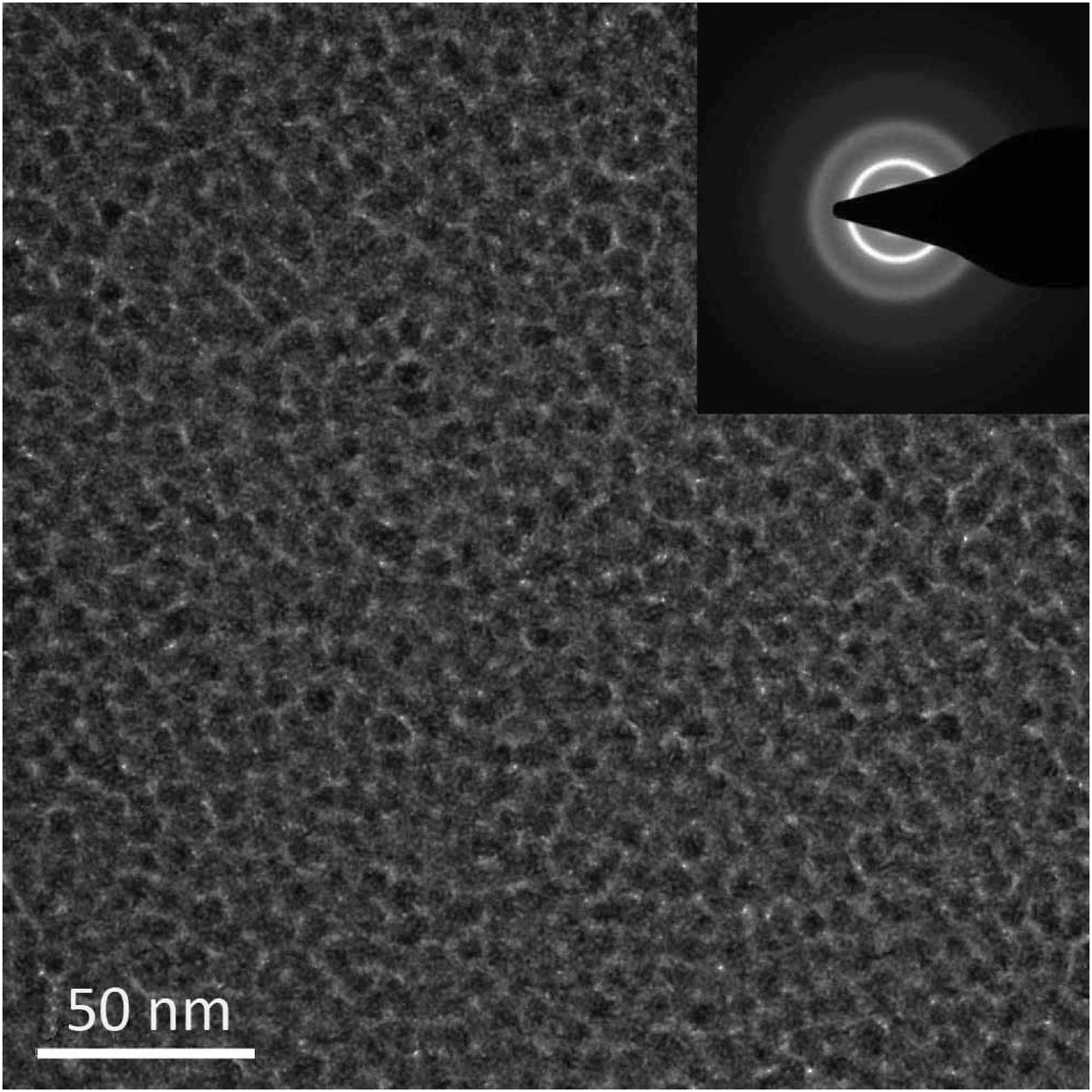}%
\caption{Bright-field micrograph and associated diffraction-pattern of the
as-deposited Ge$_{\text{x}}$Te film. The lumpy appearance of the film
morphology is characteristic of many amorphous structures and it includes the
contribution of the amorphous carbon that is the substrate in this case. }%
\end{figure}
Crystalline samples of Ge$_{\text{x}}$Te were obtained from the amorphous
Ge$_{\text{x}}$Te deposits by subjecting them to~temperatures in the range of
470-490K for 2-3 minutes. The amorphous-crystalline transformation is
reflected in the optical properties of the films as a mild change in color
tint. In this regard the result is very similar to corresponding situation in
the GeSb$_{\text{x}}$Te$_{\text{y}}$ compound studied previously \cite{12}\ as
can be seen in the comparison shown in Fig.2.%
\begin{figure}[ptb]%
\centering
\includegraphics[
height=2.4465in,
width=3.039in
]%
{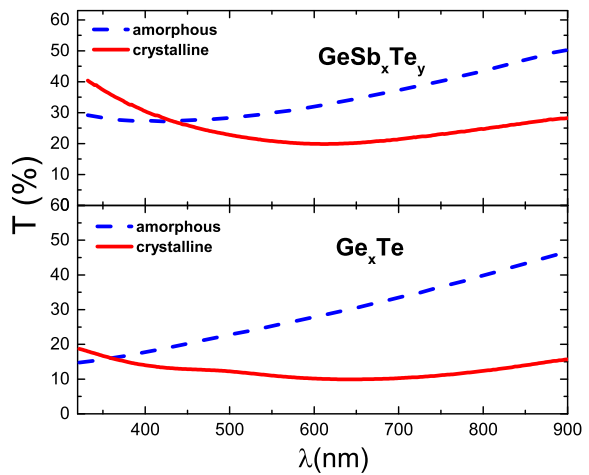}%
\caption{Optical transmission through a 50\AA ~Ge$_{\text{x}}$Te film
deposited on a 1mm glass-slide compared with that of a
120\AA \ GeSb$_{\text{x}}$Te$_{\text{y}}$ film \cite{13}.}%
\end{figure}

In terms of other properties however, there are significant differences
between our crystalline versions of Ge$_{\text{x}}$Te and the GeSb$_{\text{x}%
}$Te$_{\text{y}}$. In particular, their microstructure is different; while
both systems exhibit mosaic film structure with a tight, space-filling packing
of the crystallites, the Ge$_{\text{x}}$Te film shows a much more pronounced
preferred orientation extending over large scales. This may be seen in both
transmission electron microscope (TEM) micrographs depicted in Fig.3 (same
sample as shown in Fig.1 after crystallization at 485K) and Fig.4 (same sample
after being `aged' for a week). The diffraction patterns in these figures were
taken in selected-area mode covering 0.8~micron circle diameter. Pronounced
preferred orientation was still conspicuous using a selected-area of
4~microns, which is at least order of magnitude larger than the average size
of the grains in the studied films. This extensive texture, extending over a
scale much larger than a typical grain size, was uniformly observed across the
3 mm TEM grid by scanning it with a constant electron-beam.
\begin{figure}[ptb]%
\centering
\includegraphics[
height=3.039in,
width=3.039in
]%
{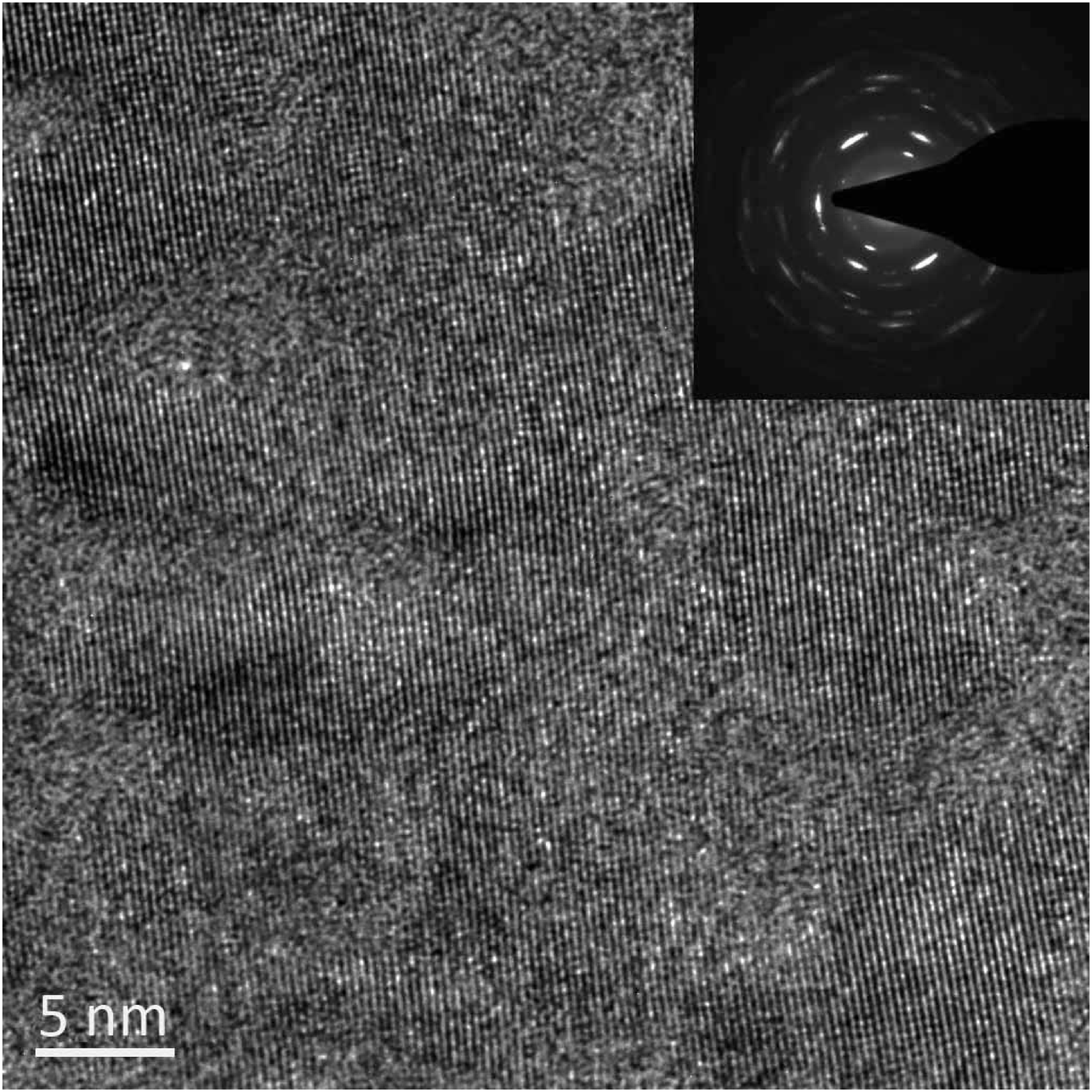}%
\caption{Bright-field micrograph and associated diffraction-pattern of the
sample in Fig.1 after crystallization. The extensive texture is clearly
observed in both the image and the diffraction pattern despite the plethora of
point defects. The diffraction pattern is consistent with the rhombohedral
(R-3m) phase of GeTe.}%
\end{figure}

We found it hard to get films with appreciable sheet resistance even in quite
thin specimen. The reason for that is presumably the reduced grain-boundary
scattering and better mobility relative to that observed in the
GeSb$_{\text{x}}$Te$_{\text{y}}$ alloys (assuming that impurity-contents, and
carrier-concentration are the same). Special measures had to be taken in
fabricating films with high values for R$_{\square}$ (which were required for
observing electron-glass properties). These included reducing the film
thickness (down to 30\AA \ relative to the constant 120\AA \ used in the study
of GeSb$_{\text{x}}$Te$_{\text{y}}$ \cite{13}), and aging the films in the lab
atmosphere. A micrograph and associated diffraction of an aged film is shown
in Fig.4.%
\begin{figure}[ptb]%
\centering
\includegraphics[
height=3.039in,
width=3.039in
]%
{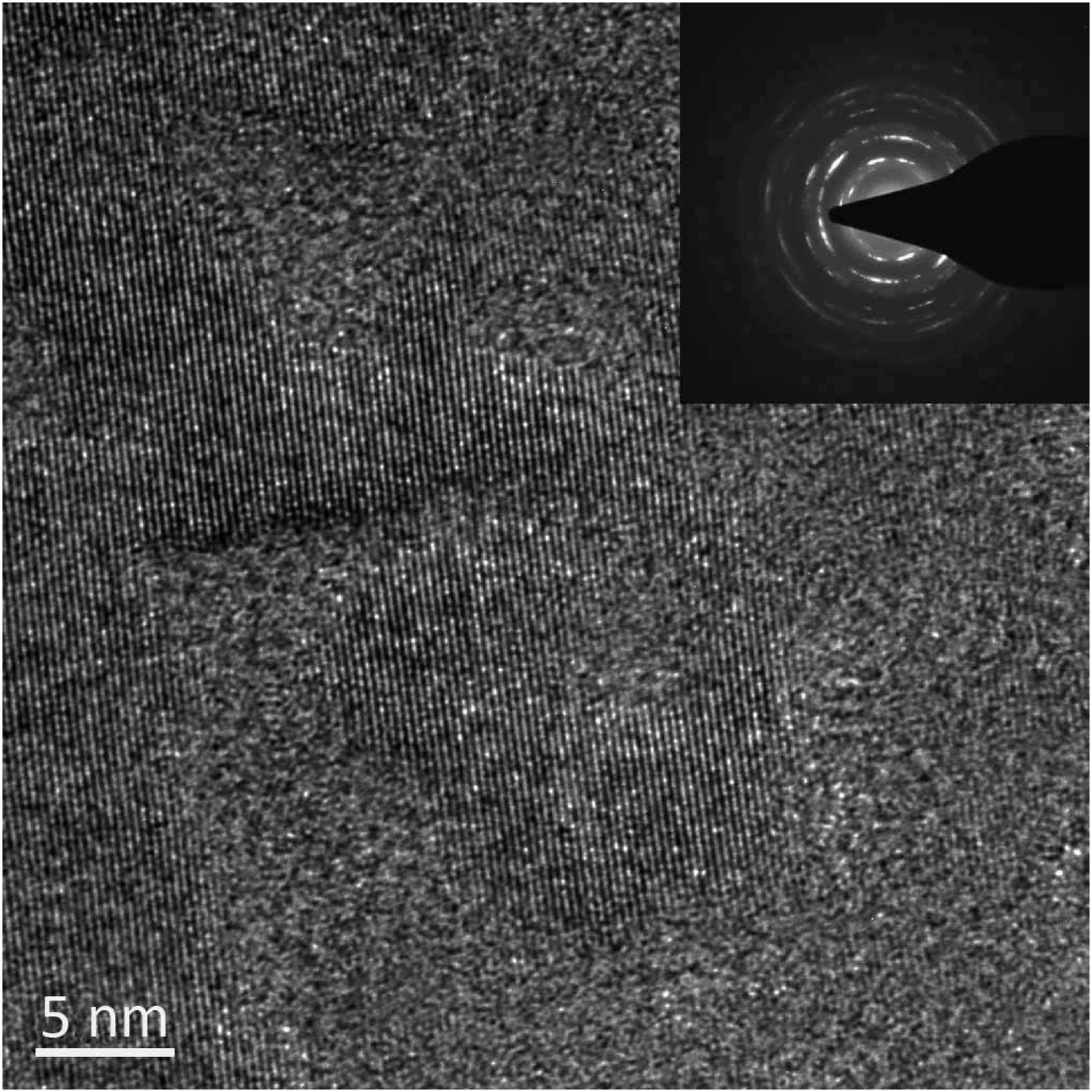}%
\caption{Bright-field micrograph and associated diffraction-pattern of the
sample in Fig.3 after aging it at room-temperature for a week (see text). Note
the increase in diffuse scattering relative to the fresh film. However, the
preferred-orientation, extending across grain boundary is still intact.}%
\end{figure}

The room-temperature resistance of our films span the range 600$\Omega
$-29k$\Omega$~which yielded 0.8k$\Omega$-5.6M$\Omega$ at T$\approx$4K. The
upper limit of this range was obtained by reducing the film thickness to
30\AA ~and exposing the sample to the lab atmosphere for several days.
However, we were yet unable to produce films with R$_{\square}$%
$>$%
5.6M$\Omega$ at 4.1K. By comparison, using the GeSb$_{\text{x}}$Te$_{\text{y}%
}$ alloy it was rather easy to obtain samples with R$_{\square}$ as high as
50M$\Omega$ even with films that were 3-4 thicker \cite{12} than the
Ge$_{\text{x}}$Te used in this work.

The diffraction pattern of the aged film shown in Fig.4 exhibited an increase
in the diffuse scattering (compare Fig.4 with Fig.3) suggestive of an
increased disorder. This is possibly due to enhanced surface scattering or the
creation of a disordered dead-layer by oxidation process at the film-air
\ boundary. Energy dispersive spectroscopy revealed a 10-15\% increase in
oxygen content in the aged film relative to the fresh sample without a
noticeable change in the Ge/Te ratio (being $\approx$0.9$\pm$0.05). X-ray
photoelectron spectroscopy on GeTe pellets exposed to lab atmosphere found
adventitious layers of carbon on their surface, which may also be source of
enhanced scattering in films exposed to air. Another difference is the
relative intensity of some rings that became more noticeable in the aged
sample. Selected-area diffraction, sampling a circle of 0.8$\mu$m diameter was
used to scan across the sample and the same features were observed throughout,
so structural inhomogeneities in this material are probably limited to spatial
scales $\leq$1$\mu$m.

The Hall effect that was monitored for some of the films showed a rather small
change during the aging process; for a $\approx$100\% increase of the film
R$_{\square}$ the Hall resistance has increased by $\approx$6\%. Based on
these Hall effect measurements the carrier-concentration n of the films was
\textit{N}=(1.5-2)x10$^{\text{20}}$cm$^{\text{-3}}$, somewhat smaller than the
\textit{N}=(4-9)x10$^{\text{20}}$cm$^{\text{-3}}$ found in the GeSb$_{\text{x}%
}$Te$_{\text{y}}$ compound \cite{13}. In both cases the Hall effect had the
sign of p-type carrier consistent with theoretical prediction for the material
\cite{18}. The latter, based on equilibrium concentration of Ge vacancies in
the ideal crystal anticipated carrier-concentration of $\approx$%
10$^{\text{19}}$cm$^{\text{-3}}$ holes. The carrier-concentration in our
films, larger by roughly order of magnitude is probably a result of the
abundant structural defects (readily observable in the TEM micrographs, Fig.3
and Fig.4) that apparently allows more Ge vacancies than the ordered crystal
can sustain in equilibrium.

\subsection{Measurement techniques}

Conductivity of the samples was measured using a two terminal ac technique
employing a 1211-ITHACO current preamplifier and a PAR-124A lock-in amplifier.
All measurements were performed with the samples immersed in liquid helium at
T$\approx$4.1K held by a 100 liters storage-dewar. This allowed up to two
months measurements on a given sample while keeping it cold (and in the dark)
which was utilized to extend the time-duration of relaxation processes as well
as many cycles of excitation-relaxation experiments. Fuller measurement
techniques related to electron-glass properties are described elsewhere
\cite{19}.

The ac voltage bias in conductivity measurements was small enough to ensure
near-ohmic conditions (except for the current-voltage plots and the `stress
protocol' described in the Results section below). Optical excitations in this
work were accomplished by exposing the sample to an AlGaAs diode operating at
$\approx$0.88$\pm$0.05$\mu$m, mounted on the sample-stage typically $\approx
$10-15mm from the sample. The diode was energized by a computer-controlled
current-source (Keithley 220).

\section{Results and discussion}

\subsection{Persistent photo-conductivity in Ge$_{\text{x}}$Te}

A main difference between the transport properties of Ge$_{\text{x}}$Te and
the GeSb$_{\text{x}}$Te$_{\text{y}}$ alloy is their different sensitivity to
optical excitation. The experimental protocol used for observing
photoconductivity is illustrated in Fig.5 using a diffusive Ge$_{\text{x}}$Te
film with R$_{\square}$=5k$\Omega$ and, for comparison, a GeSb$_{\text{x}}%
$Te$_{\text{y}}$ film with similar R$_{\square}$ and thickness measured under
the same conditions. The experiment begins $\approx$24 hours after the sample
is cooled-down to 4.1K by recording G(t) for 1-2 minutes to establish a
baseline conductance G$_{0}$. The IR source is then turned on for 3 seconds
then turned off while G(t) continues to be measured. The brief IR burst causes
G to promptly increase by $\delta$G$_{\text{IR}}$ which decays slowly with
time once the source is turned off (Fig.5). Both samples exhibit excess
conductance that persists for a long time after the optical excitation. In
terms of magnitude, the persistent photoconductivity (PPC) signal is however
much more conspicuous in the GeSb$_{\text{x}}$Te$_{\text{y}}$ film at all
values of R$_{\square}$. A detailed comparison of the PPC magnitude versus
R$_{\square}$ illustrating the difference between the two systems is given in
Fig.6. This figure includes three GeMn$_{\text{x}}$Te$_{\text{y}}$ samples.
These were prepared by co-depositing Mn with the Ge$_{\text{x}}$Te compound to
test the effect of magnetic impurities. The Mn inclusion had only a small
effect on the samples mobility, reducing it by 10-20\% (for $\approx$20\% Mn)
relative to the pure compound. As shown in Fig.6 it also had a negligible
effect on the PPC performance of the compound.
\begin{figure}[ptb]%
\centering
\includegraphics[
height=2.4396in,
width=3.039in
]%
{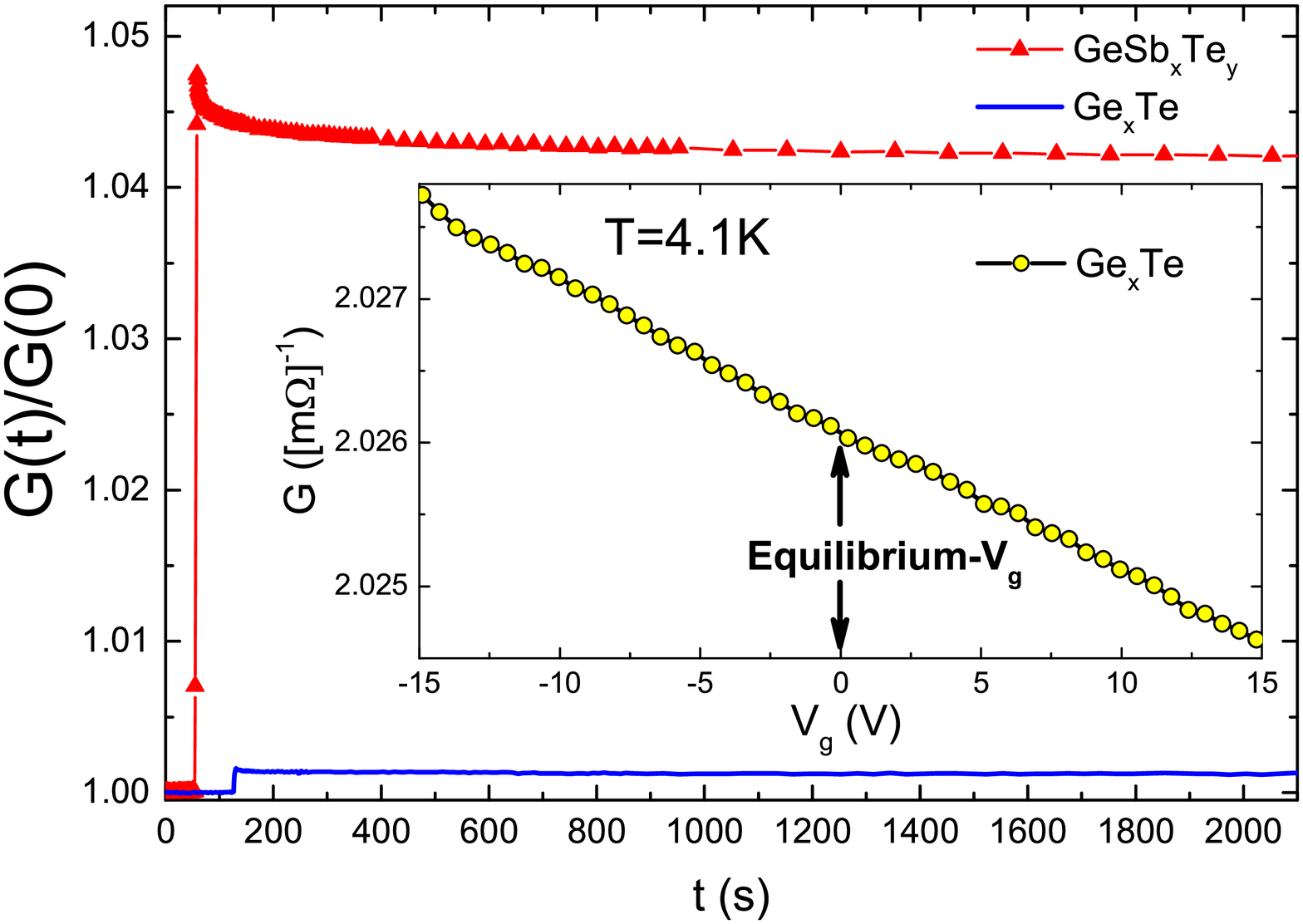}%
\caption{Comparing the persistent-photoconductivity effect typically observed
in our Ge$_{\text{x}}$Te samples with that of a GeSb$_{\text{x}}$%
Te$_{\text{y}}$ film. The R$_{\square}$ of the Ge$_{\text{x}}$Te film used
here is 5k$\Omega$ while R$_{\square}$ of the GeSb$_{\text{x}}$Te$_{\text{y}}$
film is 3k$\Omega$. The same protocol was used for both samples; same infrared
intensity, distance from film, and duration of exposure. Inset shows the
field-effect for the Ge$_{\text{x}}$Te film.}%
\end{figure}
\begin{figure}[ptb]%
\centering
\includegraphics[
height=2.4396in,
width=3.039in
]%
{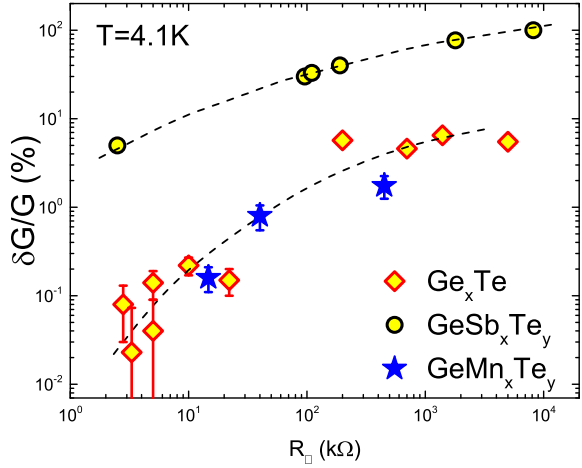}%
\caption{The relative magnitude of the infrared induced excess conductance in
the PPC state for our Ge$_{\text{x}}$Te films as function of their
R$_{\square}$. These data are compared with the respective data for the
GeSb$_{\text{x}}$Te$_{\text{y}}$ studied previously \cite{13} and measured
under the same conditions. Also shown are results for three Mn-doped
Ge$_{\text{x}}$Te samples. Dashed lines are guides to the eye.}%
\end{figure}

Actually it appears that the PPC in pure Ge$_{\text{x}}$Te and GeMn$_{\text{x}%
}$Te$_{\text{y}}$ compound differs from that in the GeSb$_{\text{x}}%
$Te$_{\text{y}}$ system not just by magnitude. The relaxation law that fits
the time dependence of the excess conductance $\delta$G$_{\text{IR}}$ in the
GeSb$_{\text{x}}$Te$_{\text{y}}$ compounds showed a rather good fit to a
stretched exponential law: $\delta$G$_{\text{IR}}$(t)$\varpropto\exp
$\{-(t/$\tau$)$^{\beta}$\} with $\beta$=0.1 for all samples with R$_{\square}$
in the 10$^{\text{3}}$-10$^{\text{7}}\Omega$ range. A similar expression could
be fitted to the PPC data of our most resistive Ge$_{\text{x}}$Te films
($\delta$G$_{\text{IR}}$ for the lower resistance samples was too small to
allow a meaningful fit) but with $\beta$=0.14-0.22. An example of a fit is
shown in Fig.7.%
\begin{figure}[ptb]%
\centering
\includegraphics[
height=2.3263in,
width=3.039in
]%
{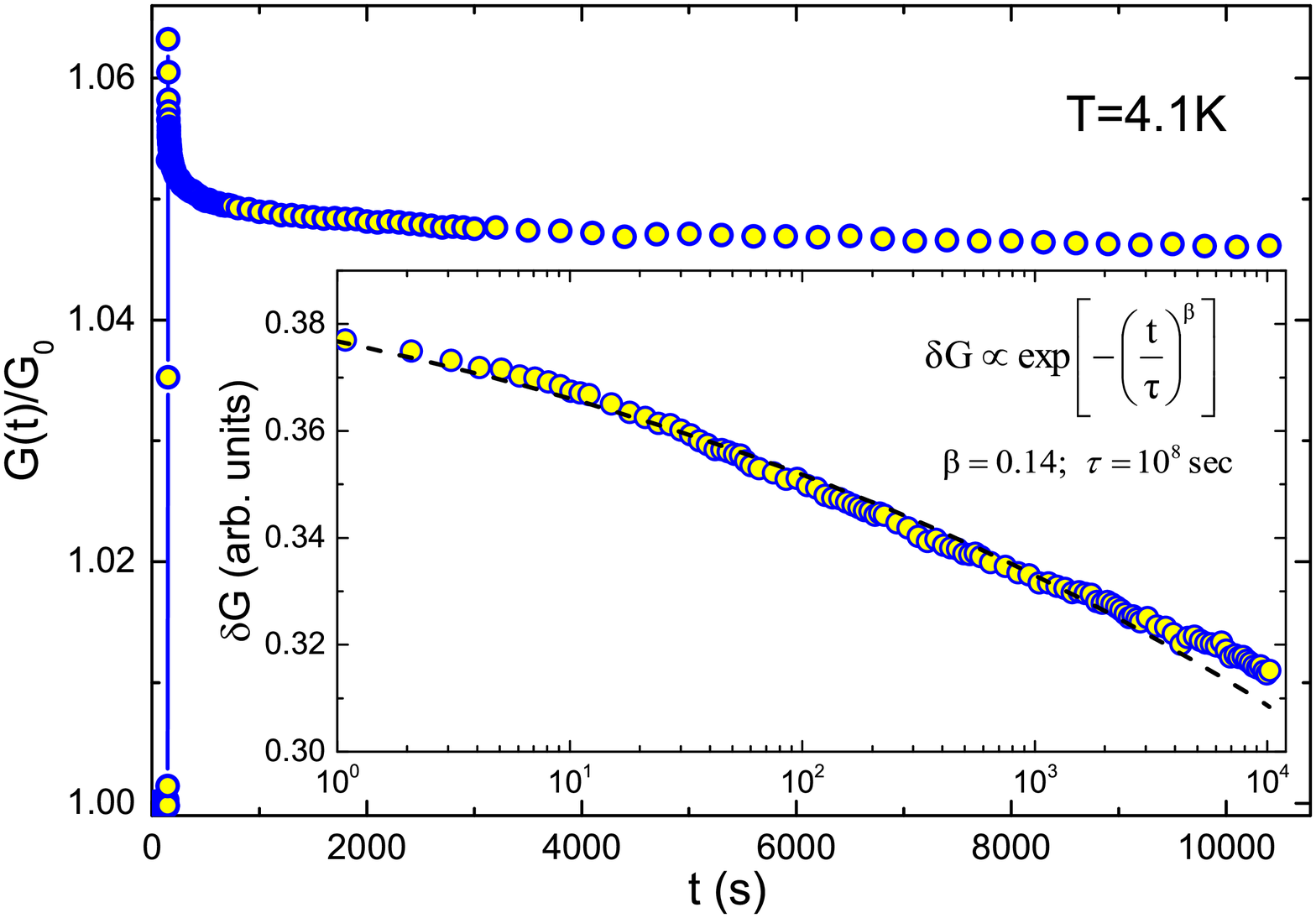}%
\caption{The persistent photoconductivity of a Ge$_{\text{x}}$Te film with
R$_{\square}$=750k$\Omega$. The inset depicts a fit (dashed line) to a
stretched-exponent for the associated excess conductance as function of time.}%
\end{figure}

This, and the much smaller $\delta$G$_{\text{IR}}$ (all other things being
equal) suggests that the presence of the Sb plays a similar role in enhancing
the PPC performance in germanium-tellurides compounds as that of In impurities
in lead-telluride alloys \cite{19}. It would be of interest to find what other
elements are effective in enhancing PPC in these systems. That is important
for understanding the basics of the PPC phenomenon but also as a tool for
elucidating the physics of the electron-glass; the PPC, when prominent enough,
may be an effective way to increase the carrier-concentration in a system
which is an important parameter in controlling the dynamics of the
electron-glass \cite{21}.

\subsection{Strongly localized Ge$_{\text{x}}$Te is an intrinsic
electron-glass}

Like in previously studied materials, a pre-requisite for observing
electron-glass features is that the system must be strongly-localized. This
applies in particular to the appearance of a memory-dip (MD) in the
field-effect measurement, which is the identifying signature of the intrinsic
\cite{13} electron-glass. A memory-dip appeared in our films at T$\approx$4K
once their R$_{\square}\gg\hslash$/e$^{\text{2}}$. A well developed MD can be
seen in Fig.8 for a Ge$_{\text{x}}$Te film with R$_{\square}$=195k$\Omega$.

The sign of $\frac{\partial}{\partial V_{\text{g}}}$G(V$_{\text{g}}$)
(reflecting how thermodynamic density of states $\partial$n$/\partial\mu$
changes with energy) is consistent with hole conduction (Fig.2 and Fig.8)%
\begin{figure}[ptb]%
\centering
\includegraphics[
height=2.3263in,
width=3.039in
]%
{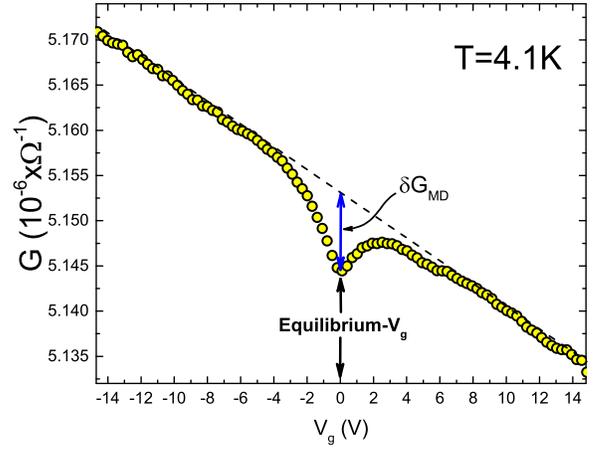}%
\caption{The field-effect for a Ge$_{\text{x}}$Te sample with R$_{\square}%
$=195k$\Omega$ showing a a memory-dip with a relative magnitude of $\approx
$0.75\% magnitude (defined as by $\delta$G$_{\text{MD}}$/Geq where Geq is the
equilibrium value of the conductance at the bath temperature). The dashed line
is the thermodynamic part of the field-effect measurement (as in the
weakly-disordered sample in Fig.5).}%
\end{figure}
and the sign of Hall-effect measurements on these films. Both the slope of
G(V$_{\text{g}}$) and the relative magnitude of the MD increases with disorder
as shown in Fig.9. The disappearance of the MD as the system approaches the
diffusive regime is common to \textit{all} intrinsic electron-glasses, and has
been seen in both two-dimensional and three-dimensional systems \cite{22}.
This is a crucial attribute of the phenomenon and should be the starting point
for any theoretical model.%
\begin{figure}[ptb]%
\centering
\includegraphics[
height=2.2701in,
width=3.039in
]%
{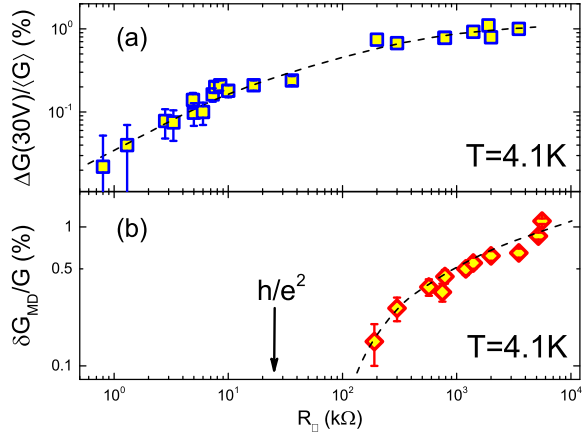}%
\caption{(a) The relative change of conductance of the antisymmetric part of
the field-effect as function of the Ge$_{\text{x}}$Te films R$_{\square}$. It
is defined as: G(-30V)/G(+30V)-1. (b) The relative magnitude of the memory-dip
for the GeTe films as a function of their R$_{\square}$. Dashed lines are
guides to the eye.}%
\end{figure}

As may be expected, Ge$_{\text{x}}$Te films that exhibit MD also show the
other characteristic features of electron-glasses. Fig.10 shows the excitation
and ensuing relaxation of the excess conductance due to a sudden change of the
gate voltage.%
\begin{figure}[ptb]%
\centering
\includegraphics[
height=2.1075in,
width=3.039in
]%
{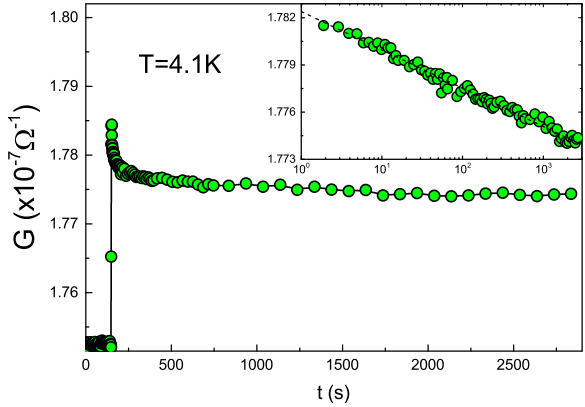}%
\caption{Conductance versus time describing excitation of the electron-glass
by a sudden change of the gate voltage. Sample has R$_{\square}$=5.6M$\Omega$.
The protocol involved a fast sweep of V$_{\text{g}}\ $(within 2s) from the
equilibrium value V$_{\text{g}}$=0V, where the system spent 24 hours, to
V$_{\text{g}}$=-10V where it was left for the duration of the time shown. The
inset illustrates the logarithmic relaxation law of the excess conductance
produced by the V$_{\text{g}}$ change.}%
\end{figure}

Another way to take the system away from equilibrium is the `stress-protocol'
\cite{23}. By applying sufficiently strong electric field across the sample
(between the `source' and `drain'), the system accumulates energy in excess of
its thermal energy. This translates into excess conductance $\Delta$G building
up over the time the field is on. Once the strong field is removed, and the
conductance is monitored under Ohmic conditions, $\Delta$G decays with time
and G approaches its equilibrium value. Both the buildup and decay of $\Delta
$G involve a protracted process. Unlike the sudden increase of G when
V$_{\text{g}}~$is$~$switched (Fig.10), $\Delta$G grows continuously throughout
the stress period without saturating. This is the analogue of the
`time-dependent heat-capacity' typical of glasses \cite{24} which is due to
the wide temporal spectrum of the system degrees of freedom. The stress
protocol is illustrated in Fig.11a and Fig.11b for the conductance evolution
G(t) during the relaxation and during the stress respectively. The conductance
dependence on the applied voltage of this sample is shown in Fig.12 with the
voltage values used during the stress and relaxation periods are marked on the
G(V) curve.
\begin{figure}[ptb]%
\centering
\includegraphics[
height=2.2286in,
width=3.039in
]%
{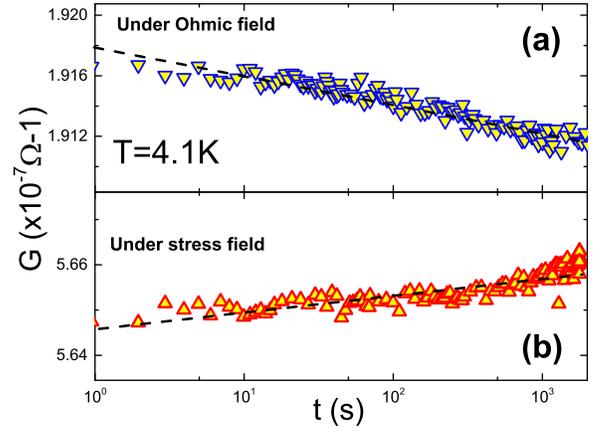}%
\caption{The two parts of the stress protocol used on a Ge$_{\text{x}}$Te
sample with R$_{\square}$=5.4M$\Omega.$ (a) Relaxation of the excess
conductance after the stress was relieved. (b) The slow conductance buildup
during the time stress was on.}%
\end{figure}

The results of the stress-protocol (Fig.11) are essentially the same in all
previously studied systems that exhibit electron-glass attributes \cite{25}.
Qualitatively different behavior has been observed in granular systems
\cite{26}. Granular systems differ from Anderson insulators in other aspects
as well although they share some glassy features like a memory-dip \cite{27}.

It is also worth commenting on the mechanism by which the applied non-ohmic
field takes the system out of equilibrium. The enhanced conductance that
appears immediately after applying the field is associated with adiabatic
modification of the hopping probabilities \cite{28}. This is the dominant
effect when the resistance is large but Joule-heating is to some degree also
responsible to the increase of G when a large voltage is applied (except when
the applied field frequency exceeds the electron-phonon inelastic-rate
\cite{29}). Joule-heating is the reason for the slow buildup of excess
conductance observed under large voltages. Qualitatively similar behavior is
achieved by raising the bath temperature. However, comparing the behavior of
G(t) under field F versus that of raising the bath temperature by $\Delta$T
(to achieve the same initial $\Delta$G), demonstrated that under $\Delta$T the
ensuing excess conductance increased with time at a faster rate \cite{30}.
This is just a manifestation of the fact alluded to above; heating is only
part of the reason for non-ohmicity in the hopping regime. The advantage of
using voltage-swings over raising-lowering the bath-temperature is the higher
speed and controllability of the procedure. The price is the uncertainty in
assigning a value of "effective-temperature" to the stress protocol; the value
of G under non-ohmic fields is, in general, not a reliable thermometer;
nonohmic measurements of G(V) are \textit{not} simply related to the
\textit{equilibrium} values of G(T).%
\begin{figure}[ptb]%
\centering
\includegraphics[
height=2.2312in,
width=3.039in
]%
{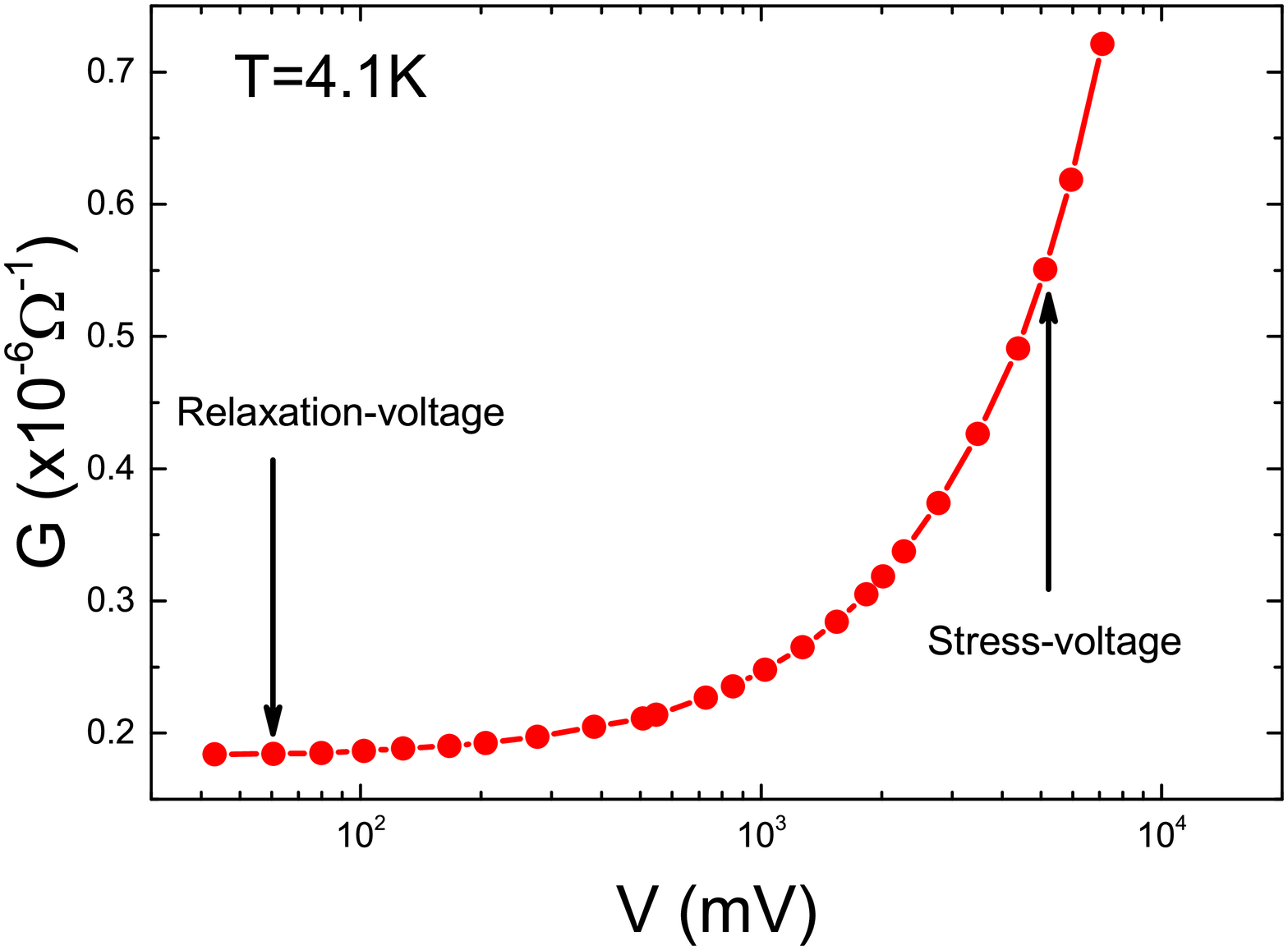}%
\caption{The conductance of the sample used for the stress protocol (Fig.11)
as function of the applied voltage. Sample dimensions are L=W=0.5mm. Marked
\ by arrows are the voltages used during the stress and relaxation periods.}%
\end{figure}

As mentioned above, the visibility of the memory-dip, over the antisymmetric
G(V$_{\text{g}}$) (controlled by the energy dependence of the thermodynamic
DOS), increases with R$_{\square}$. The memory-dip for one of the most
resistive samples we were able to manufacture in this study is shown in Fig.13
where it is compared with the MD of a GeSb$_{\text{x}}$Te$_{\text{y}}$ film
with the same R$_{\square}$ and measured under the same conditions
(temperature, sweep-rate, and gate-voltage range).%
\begin{figure}[ptb]%
\centering
\includegraphics[
height=2.3203in,
width=3.039in
]%
{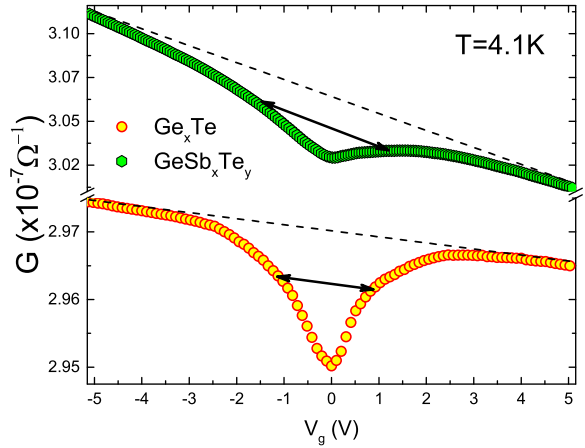}%
\caption{Comparing the memory-dip of a Ge$_{\text{x}}$Te sample with
R$_{\square}$=3M$\Omega$ with the memory-dip of a GeSb$_{\text{x}}%
$Te$_{\text{y}}$ sample with R$_{\square}$=6.2M$\Omega.$ Both G(V$_{\text{g}}%
$) curves were taken with the same sweep rate. The arrows delineate the
typical width of each memory-dip,}%
\end{figure}

The main difference between the two G(V$_{\text{g}}$) curves in Fig.13 is the
steeper antisymmetric contribution of the GeSb$_{\text{x}}$Te$_{\text{y}}$
sample. Closer examination reveals that the typical width of the MD is also
somewhat wider for the Sb-doped alloy. The narrower width of the MD in
Ge$_{\text{x}}$Te may be a result of the smaller carrier-concentration in this
material which is in line with the general trend observed in previously
studied electron-glasses. To date however, the only material where it was
possible to change the carrier-concentration over a considerable range is
amorphous indium-oxide \cite{15}. The carrier-concentration in Ge$_{\text{x}}%
$Te is associated with Ge vacancies \cite{17}, and as demonstrated by Bahl and
Chopra, the carrier-concentration\textit{ }in this system can be varied over a
decade by heat-treatment during crystallization \cite{16}. Ge$_{\text{x}}$Te
may then be another system that allows testing the relation between the
carrier-concentration, glass-dynamics, and the MD-width, by either controlling
the sample stoichiometry during deposition, alloying with foreign elements,
and thermal annealing. Future work will also focus on modifying the transport
parameters of this system by various dopants.

With the addition of the currently studied Ge$_{\text{x}}$Te, there are now
seven different Anderson-localized systems that exhibit intrinsic
electron-glass effects. The previously studied systems and their properties
were discussed elsewhere \cite{25}. The only feature common to all these
systems is having relatively large carrier-concentration;~\textit{N}$\geq
$5x10$^{\text{19}}$cm$^{\text{-3}}$. These systems have quite different
structural properties making it hard to conceive of a common defect that might
be responsible for the long relaxation times observed in their nonequilibrium
transport properties. Grain-boundaries for example, are not likely to be
relevant as their contribution to transport must be very weak in
Ge$_{\text{x}}$Te relative to other electron-glasses while the electron-glass
effects exhibited by all these systems are very similar; they all show slow
relaxation and a memory-dip. It is therefore more likely that it is the
\textit{magnitude} of the disorder rather than its specific nature that is the
important factor. This, in turn, suggests that quantifying the disorder in the
Anderson-insulating phase may be a vital step in the quest to unravel the
mechanism responsible for the electron-glass dynamics.

\begin{acknowledgments}
This research has been supported by a grant administered by the 1126/12 grant
administered by the Israel Academy for Sciences and Humanities.
\end{acknowledgments}

\end{document}